\documentclass[aps,prl,onecolumn,12pt,groupedaddress]{revtex4-1}
\usepackage{graphicx}
\begin{document}


\title{Dispersion assessment of three-dimensional Phononic Crystals using Laser Doppler Vibrometry}


\author{I. K. Tragazikis, D. A. Exarchos, P. T. Dalla, K. Dassios, and T. E. Matikas}
\affiliation{Dept. of Materials Science \& Engineering, University of Ioannina, 45110 Ioannina, Greece}

\author{I. E. Psarobas}
\email{ipsarob@phys.uoa.gr}
\affiliation{Section of Solid State Physics, National and Kapodistrian University of Athens, 15784 Athens, Greece}


\date{\today}

\begin{abstract}
The elastodynamic response of finite 3D phononic structures is analyzed by means of comparing experimental findings obtained through a laser Doppler vibrometry-based methodology
and theoretical computations performed with the Layer-Multiple-Scattering method.
The recorded frequency-gap spectrum of the phononic slabs  exhibited a good agreement of theory to experiment.
Along these lines, a newly developed technique, based on laser doppler vibrometry, has been proposed and validated for the dispersion efficiency in 3D phononic metamaterials.
\end{abstract}


\maketitle

\section{Introduction}
Wave propagation in inhomogeneous media is a problem of
wide interest due to the implications in technology and the scientific insight in understanding a large number of physical problems~\cite{physica}.
Classical wave transport in periodic media can provide the means to control light
(electromagnetic waves), sound (elastic waves) or both, with the development of
novel materials, also known as classical spectral gap materials. Such a periodic arrangement
of scatterers can obviously open up several directional spectral gaps.
When for all directions the spectral gaps overlap so that there is a forbidden range
of frequencies in which the waves cannot propagate in any direction, there is a
special type of material that exhibits an absolute frequency gap response. This
paper deals with composite materials whose elastic properties vary periodically
in space, and are also known as Phononic Crystals (PnC)~\cite{ZfK}. PnCs on the other hand
possess unique properties and exotic metamaterial features that can manifest in a wide
range of frequencies from a macroscopic point of view with infrasound and seismic waves,
to mesoscopic systems with ultasound and up to hypersound and optomechanics as well as nanoscale thermal devices~\cite{maldovan}.
All important physical phenomena associated with PnCs are completely scaled from the Hz to the THz regime
and therefore conclusions and results observed are independent of any limited frequency spectrum.

Advanced materials with such properties aim to control the propagation of elastic waves (vibrations) in various technologically exploitable ways.
Elastic wave transport in PnCs has attracted an extraordinary amount of attention over the last decades~\cite{springer}, in fields
ranging from frequency gap formation in 3D structures~\cite{sainidou} and
Anderson Localization of classical waves~\cite{anderson}, to the study of structures of more exotic
geometry~\cite{aip,spie}, negative modulus acoustic metamaterials~\cite{negative1,negative2,page} and even harvesting vibrations via PnC isolator modules~\cite{des}.

Omnidirectional frequency gaps do not appear easily in 3D solid PnCs.
The reason is that directional gaps corresponding to all  degrees of freedom
and for all directions of elastic waves do not necessarily overlap.
However, it has been found that PnCs from non-overlapping scatterers of high density in
a low density matrix (cermet topology) can function as absolute frequency filters,
regardless of directionality~\cite{ipsa01,sainidou}.
There are various methods available for the calculation of the elastic properties
of PnCs~\cite{ZfK}, such as the traditional band-structure methods, which
mainly deal with periodic, infinite, and nondissipative structures. However, in an
experiment, one deals with finite-size slabs and the measured quantities are, usually,
the transmission and reflection coefficients. Apart from that, realistic structures
are dispersive and exhibit losses. We recall that the usual band-structure
calculation proceeds with a given wave vector in order to compute the eigenfrequencies
within a wide frequency range together with the corresponding eigenmodes.
On the contrary, on-shell methods proceed differently: the frequency is fixed and
one obtains the eigenmodes of the crystal for this frequency. These methods are
ideal when dealing with dispersive materials (with or without losses). Moreover,
on-shell methods are computationally more efficient than traditional
band-structure methods~\cite{comphy}.

Technological advancements, especially 3D-printing technology, have made easier to construct PnC slabs that mimic exactly
the atom arrangement of crystalline matter so that one can really benefit from the bulk properties of 3D phononic structures.
Most of the experiments so far have dealt with 2D phononic structures, in which cases theoretical predictions were adequately verified.
In particular, the Layer-Multiple-Scattering (LMS) method~\cite{comphy,ZfK2},
a semi-analytical on-shell method with obvious advantages over pure numerical methods,
has been examined for its accuracy in a 2D system formed by a monolayer of spheres~\cite{2dexp}.
In this paper, starting from modeling a 3D PnC within the framework of LMS,
we have constructed 3D PnC specimens of different symmetries and, following a thorough investigation of their crystallographic integrity,
we were able to monitor their behavior in the ultrasonic regime with Laser Doppler Vibrometry (LDV)~\cite{LDVibe}, thus establishing an experimental technique
for observing the dispersion of 3D phononic metamaterials.

\section{Theory}

The layered multiple-scattering theory provides a framework for a unified description of wave propagation
in three-dimensional periodic structures, finite slabs of layered structures, systems with impurities, namely isolated
impurities, impurity aggregates, or randomly distributed impurities~\cite{physica}.
In particular, the LMS method~\cite{comphy} is well-documented for the elastodynamic response
of PnCs with spherical and non-spherical inclusions~\cite{Gatzounis}.
The method, based on an \emph{ab initio} multiple scattering
theory~\cite{physica}, constitutes a powerful tool for an accurate description of the elastic (acoustic) response of composite
structures comprised of a number of different layers having the same 2D periodicity in the $xy$-plane (parallel to
the layers). LMS provides the complex band structure of the infinite
crystal associated with a given crystallographic plane and also the transmission, reflection, and absorption coefficients
of an elastic wave incident at any angle on a slab of the crystal, parallel to a given plane, of finite thickness.
An advantage of the method is that it does not require periodicity in the $z$-direction (perpendicular
to the layers). In order to calculate the complex frequency band structure of the above crystal associated with
the elastic field in the manner described in Ref.~\cite{comphy}, periodic boundary conditions are imposed initially and then, for a given
angular frequency $\omega$ and reduced wave vector $\mathbf{k}_{\parallel}$, we obtain the eigenmodes of the elastic field by determining
$k_z$. The reduced wave vector $\mathbf{k}_{\parallel}$ (parallel to the crystallographic plane of stacking) and $\omega$ are given conserved
quantities. $k_z$ follows from the definition of the wave vector $\mathbf{k} = [ \mathbf{k}_{\parallel},k_z(\omega, \mathbf{k}_{\parallel})]$ of a generalized Bloch wave.

The accuracy of the computations performed herein by the LMS code~\cite{comphy} is determined by the cutoff values of the angular momentum number $\ell_{\rm max}$
coming from the spherical wave multipole expansion and the number $g_{\rm max}$ of reciprocal lattice vectors $\mathbf{g}$ used in the plane wave expansion,
necessary to incorporate Ewald summation techniques for faster convergence. In the following analysis,  a cubic stacking, viewed as a succession of (001) crystallographic planes,
has been considered for both the hexagonal close-packing and the body-centered cubic arrangements. For $\ell_{\rm max}=7$ and $g_{\rm max}=45$
we have established a convergence of less than 0.01\% and no numerical instabilities were observed.
Finally, all proper attenuation and realistic losses (experimentally measured) were taken into account, as the method incorporates into the calculations many complex dispersion behaviors~\cite{ipsa02}.

\begin{table}[t]
  \centering
    \caption{Mechanical properties of materials used, at room temperature.}
   \label{tab:table1}
    \begin{tabular}{l|c|c|c|c|c|c} 
    \hline
    \hline
      \textbf{Materials} & \textbf{Density} & \textbf{Longitudinal} & \textbf{Shear} & \textbf{Acoustic} & \textbf{Young} & \textbf{Shear}\\
      \  & \  & \textbf{speed} & \textbf{speed} & \textbf{Impedance} & \textbf{Modulus} & \textbf{Modulus}\\
      \hline
      \ \ & $g/cm^3$ & $m/s$ & $m/s$ & $10^6Kg/sm^2$ & GPa & GPa\\
      \hline
      \hline
     Air & $1.2\times10^{-3}$ & 343 & - & $4\times10^{-4}$ & $10^{-4}$ & -\\
      Paraffin & 0.9 & 2040 & 800 & 1.84 & 1.62 & 0.58\\
      Stainless steel & 7.78 & 5760 & 3160 & 44.8 & 200 & 77.9\\
      Aluminum & 2.7 & 6320 & 3130 & 17.06 & 70.76 & 26.45\\
      Cement paste & 1.97 & 3680 & 1990 & 7.25 & 17.5 & 6.8\\
      \hline
      \hline
    \end{tabular}
\end{table}

\begin{figure}
  \centering
  \includegraphics [width=0.5 \textwidth]{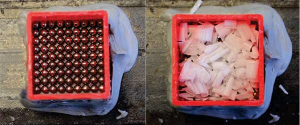}
  \caption{PnC assembly. The white material on the right is the paraffin used.}
  \label{fig01}
\end{figure}

\begin{figure}
  \centering
  \includegraphics [width=0.5 \textwidth]{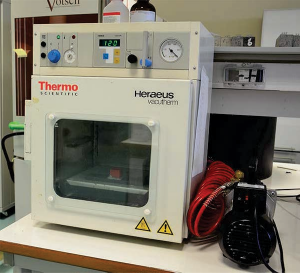}
  \caption{The vacuum oven with the PnC inside.}
  \label{fig02}
\end{figure}

\begin{figure}
  \centering
  \includegraphics [width=0.5 \textwidth]{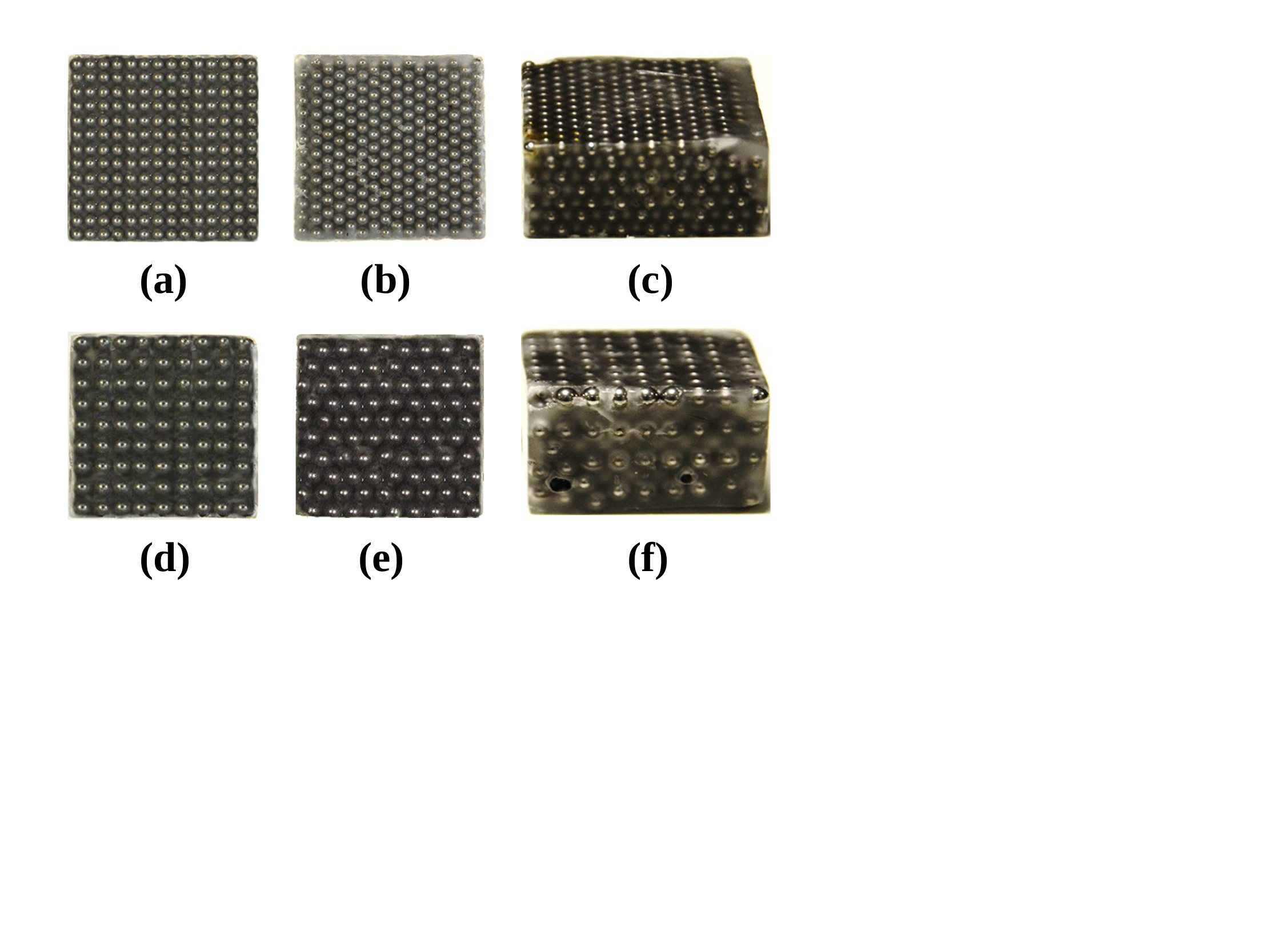}
  \caption{PnCs consisted of stainless steel spheres in paraffin matrix. The spheres have a $2\, mm$ diameter (top row) and $3\, mm$ on the bottom.
  From left to right, top layer along the (001) direction of the {\it bcc} crystal slab [(a) and (d)], top layer along the (001) direction of the {\it hcp} crystal slab [(b) and (e)].
  On the right, the 3D crystal in two different arrangements. {\it hcp} and {\it bcc} in (c) and (f), respectively.}
  \label{fig03}
\end{figure}

\begin{figure}
  \centering
  \includegraphics [width=0.5 \textwidth]{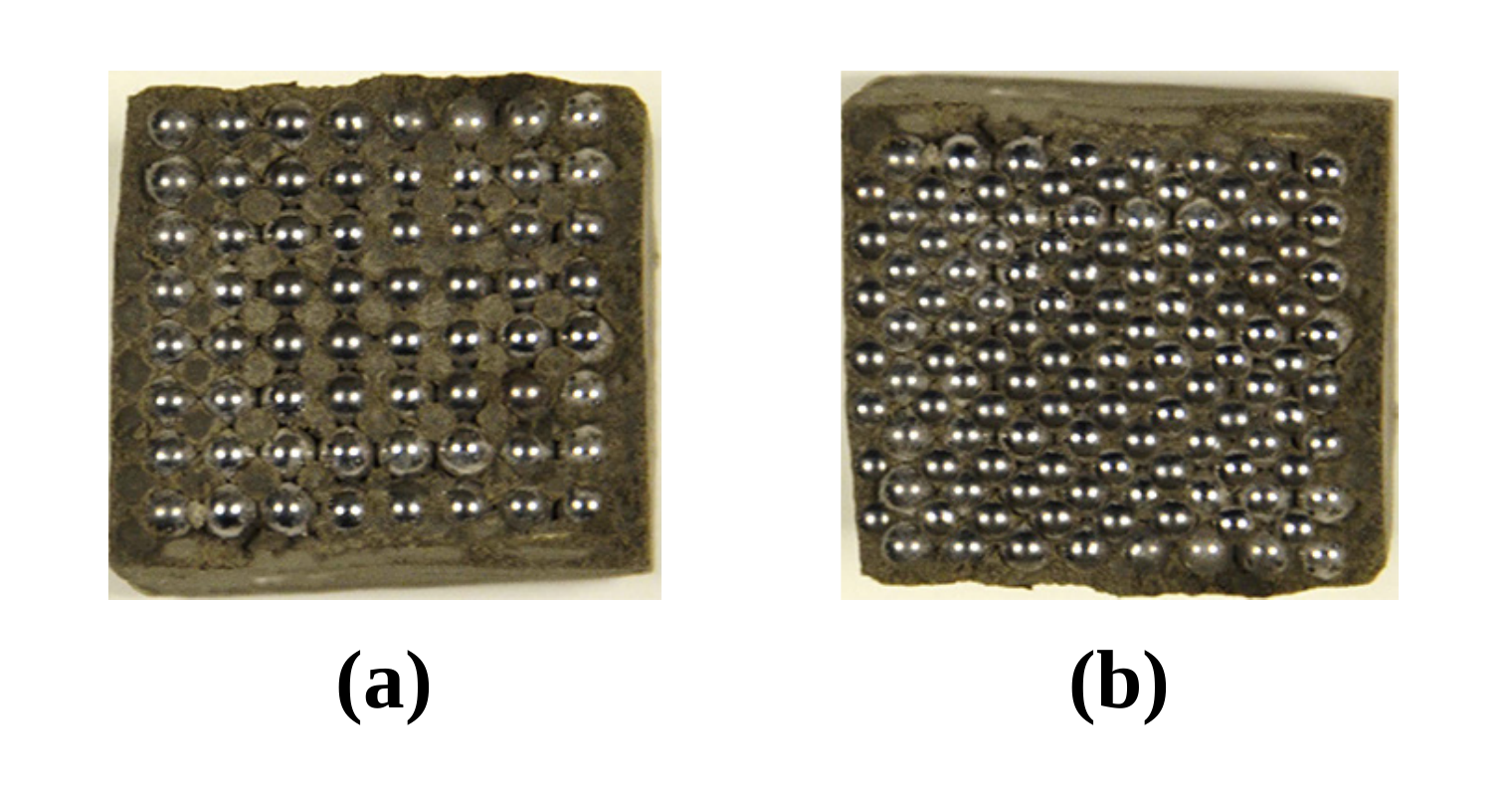}
  \caption{PnC of $3\, mm$ diameter stainless steel spheres in cement paste. (a) is a {\it bcc} crystal, while (b) is of {\it hcp} arrangement}
  \label{fig04}
\end{figure}

\begin{figure}
  \centering
  \includegraphics [width=0.5 \textwidth]{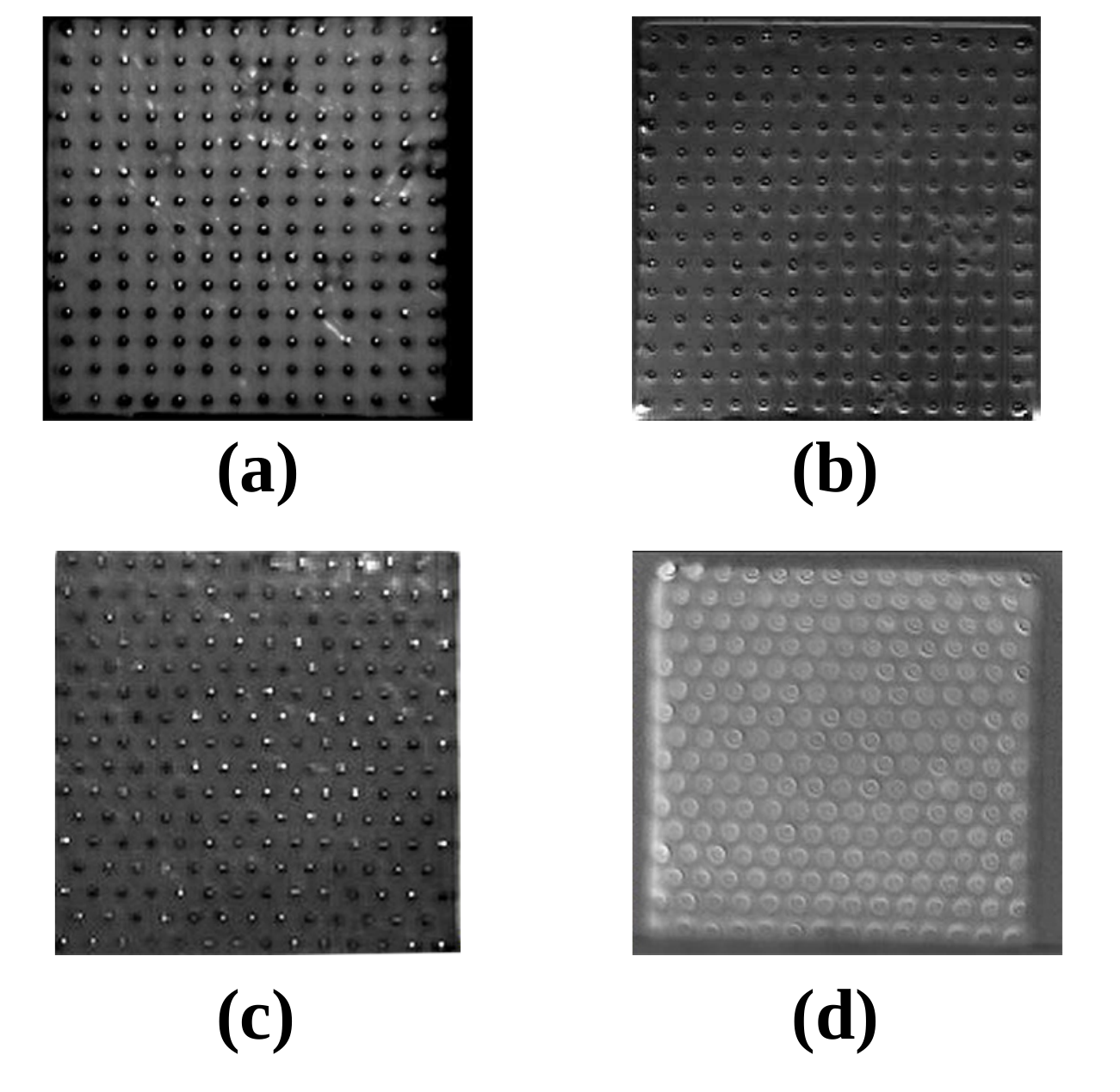}
  \caption{IR thermogram of the {\it bcc} slab of $2\,mm$ spheres in paraffin (on top), where (a) is the top PnC layer and (b) the 3rd inner layer from top.
  The bottom set corresponds to the {\it hcp} slab, where (c) is the top layer and (d) the 3rd inner layer from top.}
  \label{fig05}
\end{figure}

\begin{figure}
  \centering
  \includegraphics [width=0.5 \textwidth]{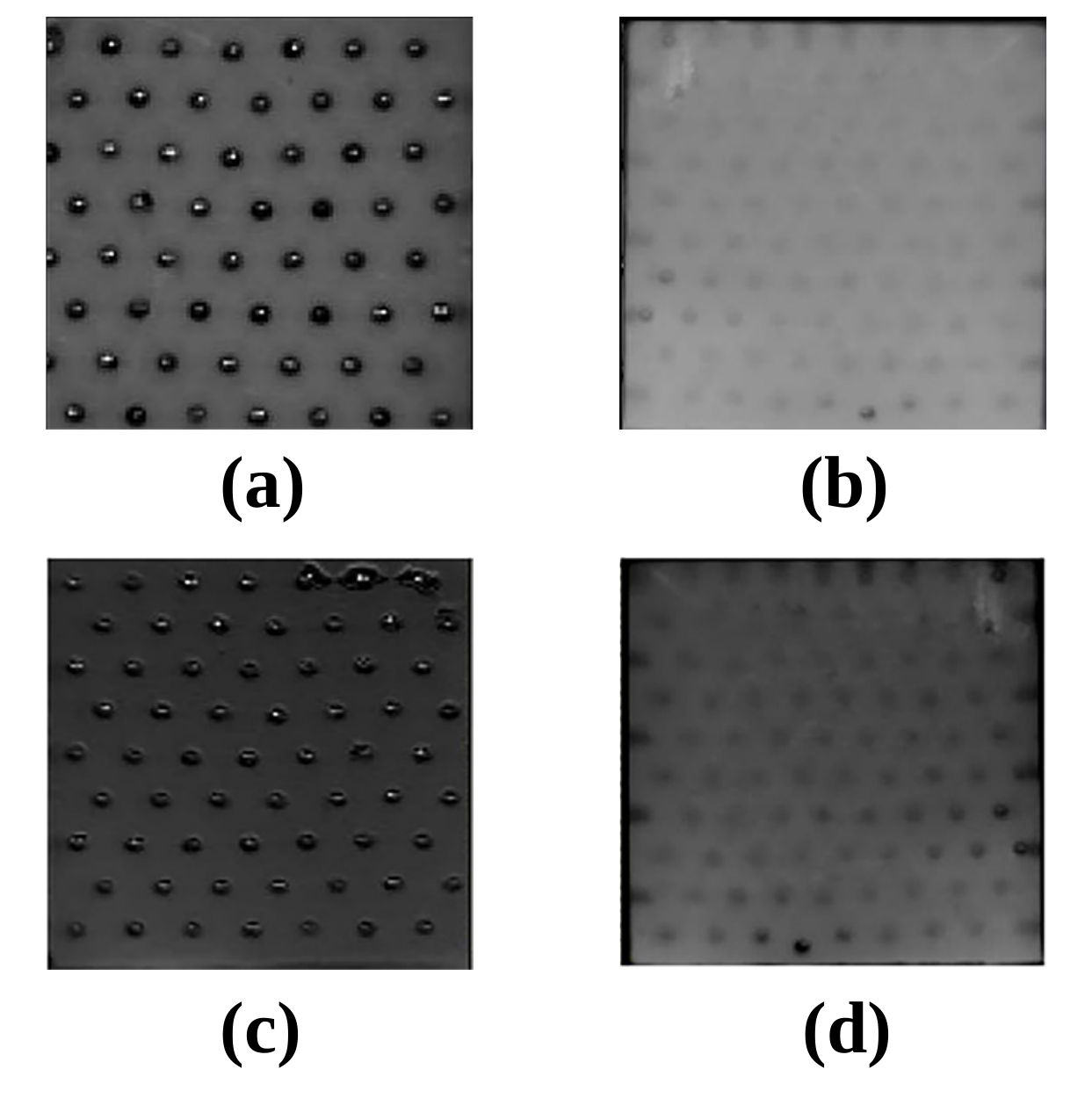}
  \caption{IR thermogram of the {\it hcp} slab of $3\,mm$ spheres in paraffin (on top), where (a) is the top PnC layer and (b) the 3rd inner layer from top.
  The bottom set corresponds to the same {\it hcp} slab in cement paste, where (c) is the top layer and (d) the 3rd inner layer from top.}
  \label{fig06}
\end{figure}

\begin{figure}
  \centering
  \includegraphics [width=0.3 \textwidth]{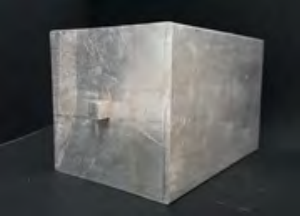}
  \caption{Aluminum waveguide. At the center of the front face of the waveguide appears a small cube of dimensions $10\times 10\times10\,  mm^3$.}
  \label{fig07}
\end{figure}

\begin{figure}
  \centering
  \includegraphics [width=0.5 \textwidth]{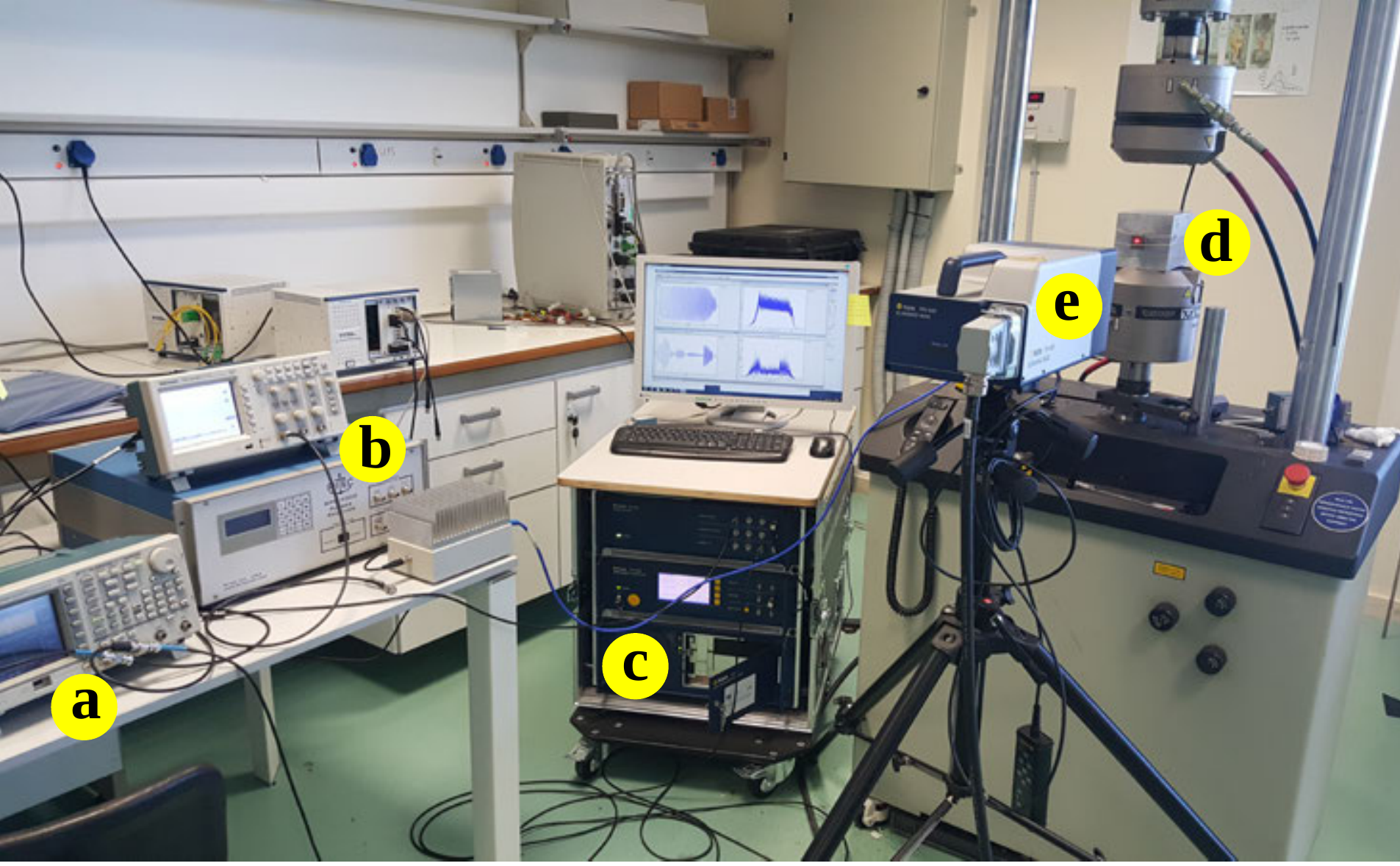}
  \caption{Experimental setup: (a) Tektronic TDS 1012B pulse generator, (b) RITEC RPR-4000 high voltage pulse generator, (c) Control unit of the Laser Doppler Vibrometer, (d) the PnC slab, and (e) the 2D scanning laser head.}
  \label{fig08}
\end{figure}

\begin{figure}
  \centering
  \includegraphics [width=0.5 \textwidth]{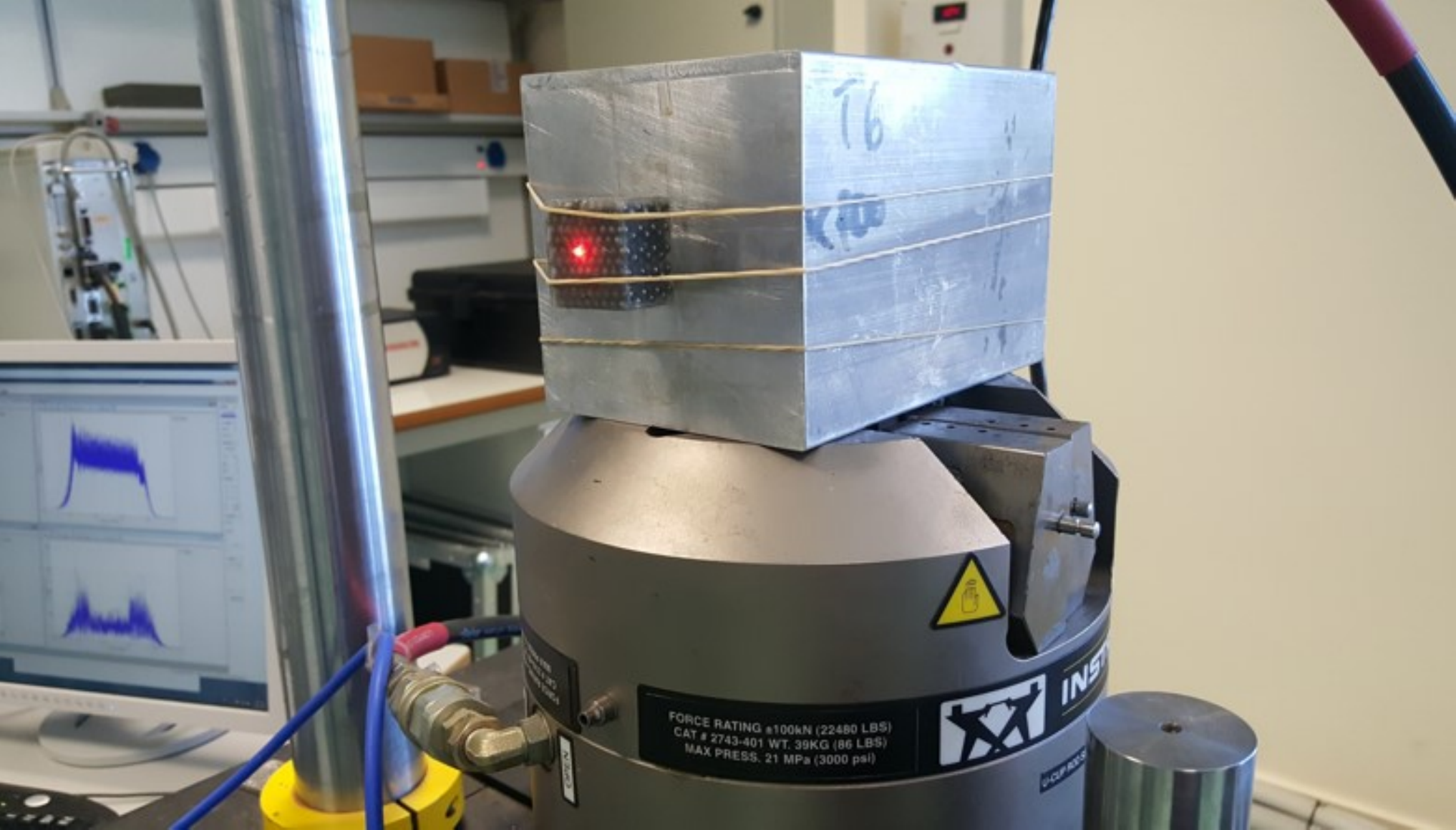}
  \caption{A detailed view of the PnC slab mounted on the waveguide. The red spot on the top of the crystal corresponds to the laser beam of the LDV.}
  \label{fig09}
\end{figure}

\section{Fabrication of Phononic Crystals}
For the construction of phononic slabs with appreciable gap width (as predicted theoretically), different small-scale phononic structures with varying volume filling fraction were used. Firstly, we have constructed slabs consisting of stainless steel spheres placed in a paraffin matrix at specific crystallographic lattice arrangements.  Optimization of phononic structures requires  a significant difference in the values of acoustic impedance between the filler and the matrix~\cite{springer}. Stainless steel was chosen as the material for the inclusions due to  its high density which enables high speeds of longitudinal waves hence endowing high acoustic resistance. Paraffin was selected as matrix owing to its small acoustic impedance. The main properties of the materials used are given in Table~\ref{tab:table1}.

Very high purity paraffin was used for the preparation of specimens with matrix transparency which facilitated visualization of the geometry of the phononic structure. Due to its very low viscosity and low surface tension, paraffin is a very good solution for the manufacture of such samples.  Stainless steel spherical inclusions of two different diameters, namely  $2\, mm$ and $3\, mm$ were used. Theoretical predictions (LMS) mandated the consideration of two crystal lattices of different symmetry, namely body centered cubic ({\it bcc}) and hexagonal close packed ({\it hcp}). Then the spheres were prepared for stacking according to each individual lattice arrangement. Before being embedded in the matrix, inclusions were cleaned in acetone and then in deionized water, in an ultrasonic bath, for removal of organic and other superficial residues originating from the production phases. Then, compressed air was used to dry the spheres which were subsequently stored in airtight glass containers to isolate inclusions from surrounding environment and avoid eventual further contamination; the containers were subjected to the same cleaning procedure. Attempts to create phononics slabs without initially cleaning the spheres failed due to low stacking efficiency and
creation of a large number of defects per level.

For the construction of phononic slabs, plastic molds of internal dimensions of $28 \times 28 \times 35\ mm^3$ were used which had been previously thoroughly cleaned. Stacking of the steel balls per level was done manually. For specimens with $2\, mm$ beads, seven-layers thick, each layer/plane consisted of 14 beads in the $x$-direction and 14 beads in the $y$-direction. Specimens with $3\, mm$-diameter beads were constructed equally thick with the difference of having 9 beads in the $x$-direction and 9 in the $y$-direction per stacking plane. The packing density of each cell in the {\it hcp} arrangement was $\sim$74\%.

After placing all layers inside the plastic matrix, very small paraffin trimmings were placed over the steel spheres as it is illustrated in Fig.~\ref{fig01}. The slab was thermally cured in a vacuum oven (Fig.~\ref{fig02}), in order to achieve uniformity. A vacuum process applied for 45 minutes in the oven environment, at room temperature, enabled complete removal of air bubbles entrapped between the steel spheres. After the end of the vacuum process,  the phononic crystal was heated to $120^{\circ}$C (paraffin melting point approximately $60^{\circ}$C) to allow reduction of paraffin viscosity and enable wetting of the steel sphere stacks. After this, the oven was allowed to cool down to room temperature, and vacuum process was released.

Based on the aforementioned procedure, phononic slabs were available for ultrasonic testing. Figures~\ref{fig03} and ~\ref{fig04} show the phononic slabs with inclusion diameters of $2\, mm$ and $3\, mm$, respectively.

Additional PnCs of equivalent quality and $3mm$ diameter inclusions were fabricated with cement paste as matrix material. Fabrication of these samples was much more demanding than their paraffin-based counterparts, due to the  higher viscosity of cement paste compared to paraffin, making its impregnation much more difficult (Fig.~\ref{fig05}).

The quality of fabricated phononic slabs and the geometric arrangement of phononic crystals in particular, was evaluated non-destructively by means of Infrared (IR) Lock-In thermography.  Infrared thermography (IRT) is the most appropriate technique for this task, as the thermal waves can penetrate both transparent materials (such as paraffin) as well as to opaque materials (such as the cement paste). As seen in Figures~\ref{fig06},~\ref{fig07},~\ref{fig08} and~\ref{fig09}, the IRT-prone geometry of both the outer and inner layers of the spheres was found to be of adequate quality.

\begin{figure}
  \centering
  \includegraphics[width=0.5 \textwidth]{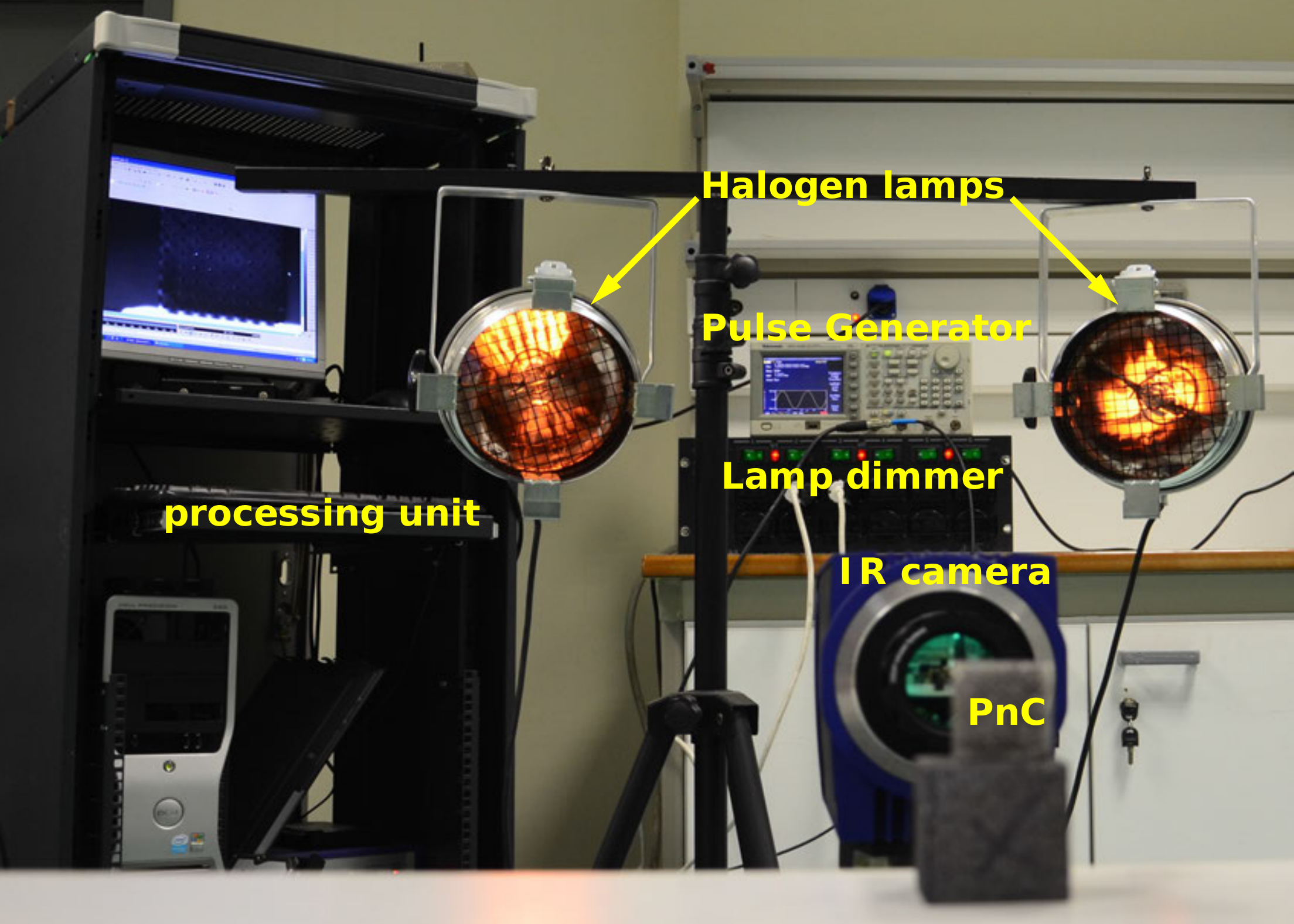}
  \caption{Setup for imaging the PnC slab using lock-in IR thermography.}
  \label{fig10}
\end{figure}

\begin{figure}
  \centering
  \includegraphics[width=0.5 \textwidth]{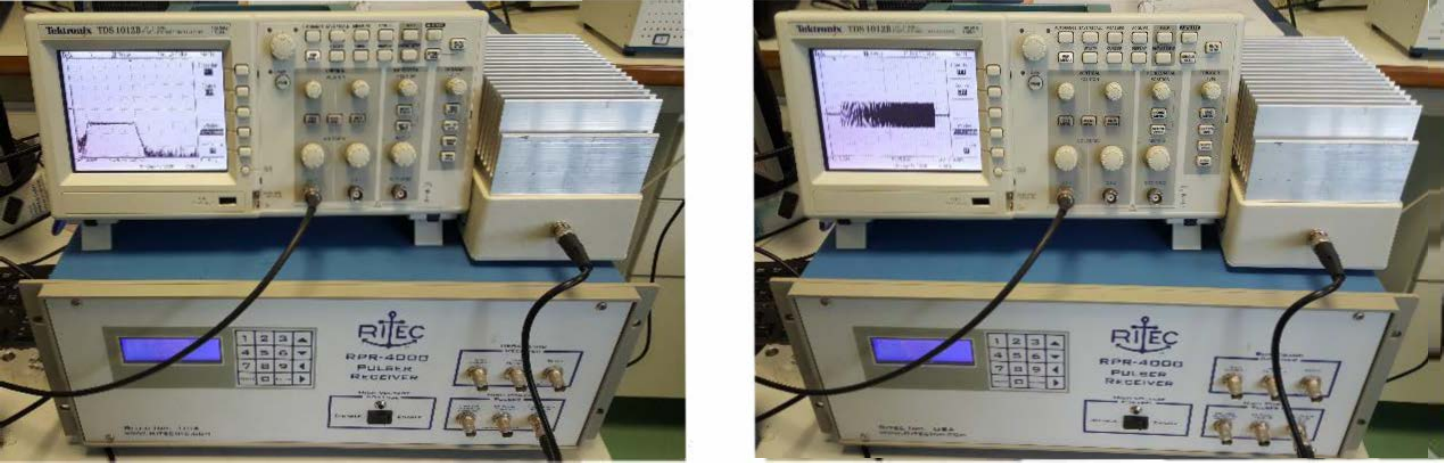}
  \caption{A chirp signal. In FFT on the left and in Time domain on the right.}
  \label{fig11}
\end{figure}

\begin{figure}
  \centering
  \includegraphics[width=0.7 \textwidth]{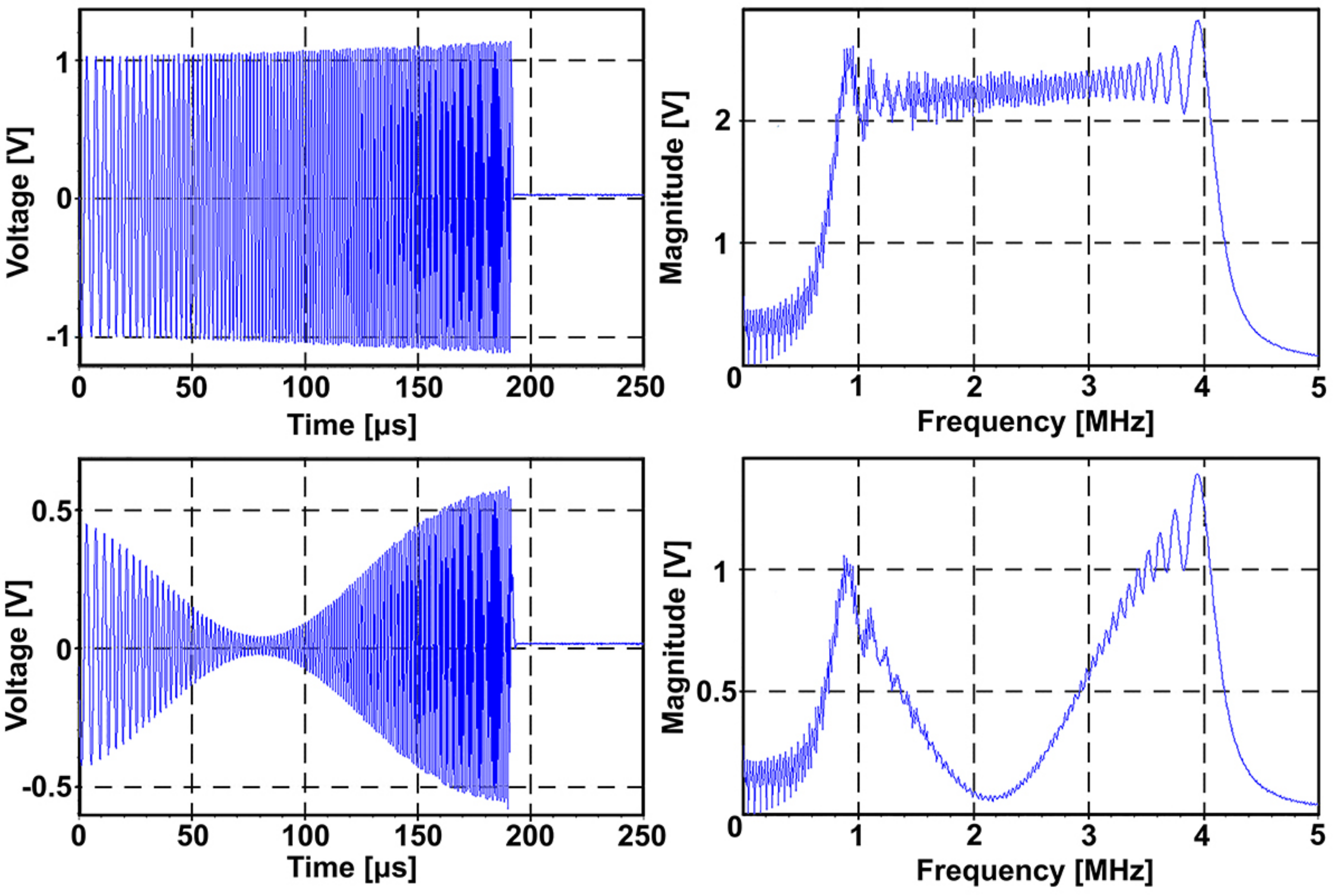}
  \caption{Time Domain chirp signal on the left and after passing through a filter (bottom left). FFT of chirp signal on the right and after passing a filter (bottom right).}
  \label{fig12}
\end{figure}

\begin{figure}
  \centering
  \includegraphics[width=0.7 \textwidth]{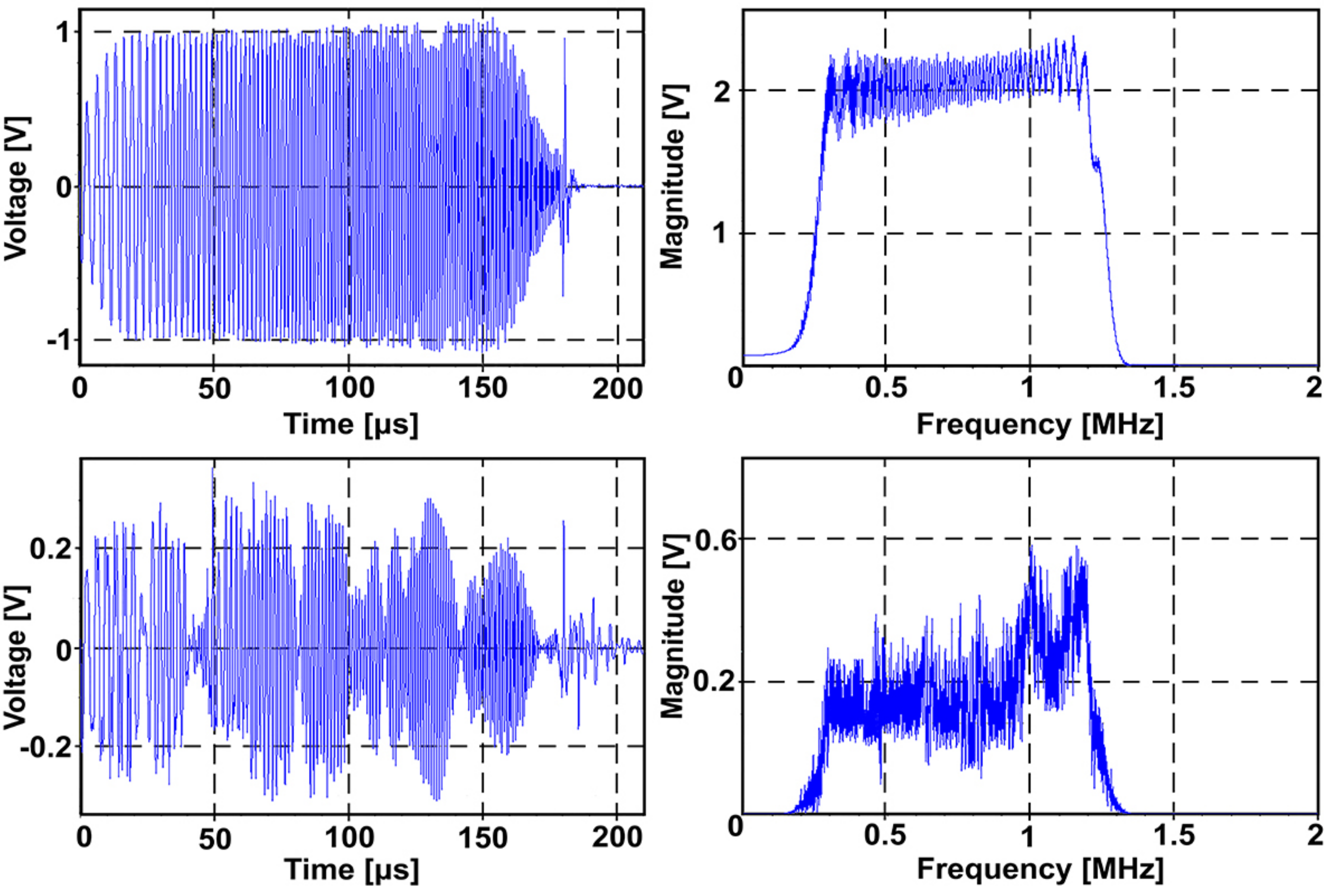}
  \caption{Experimental result of pure paraffin.}
  \label{fig13}
\end{figure}

\begin{figure}
  \centering
  \includegraphics[width=0.7 \textwidth]{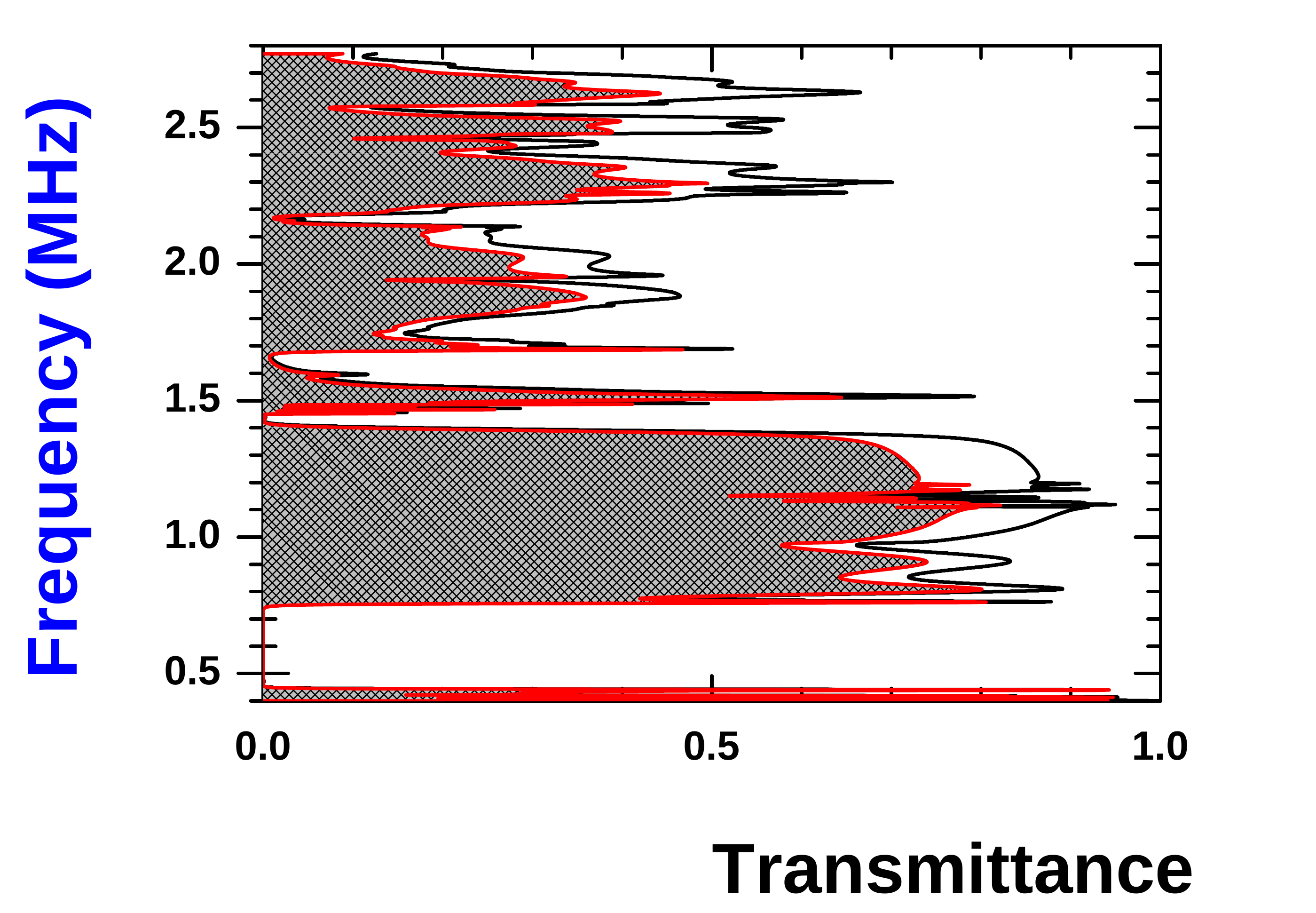}
  \caption{Calculated transmission spectrum of a longitudinal wave along the [001] direction of a {\it bcc} phononic slab of $2mm$ in diameter steel spheres in a paraffin matrix - 8 layers thick.
  The red curve denotes the case with loses. The {\it bcc} arrangement has a lattice constant ${\rm a}=2.31\,mm$.}
  \label{fig14}
\end{figure}

\begin{figure}
  \centering
  \includegraphics[width=0.7 \textwidth]{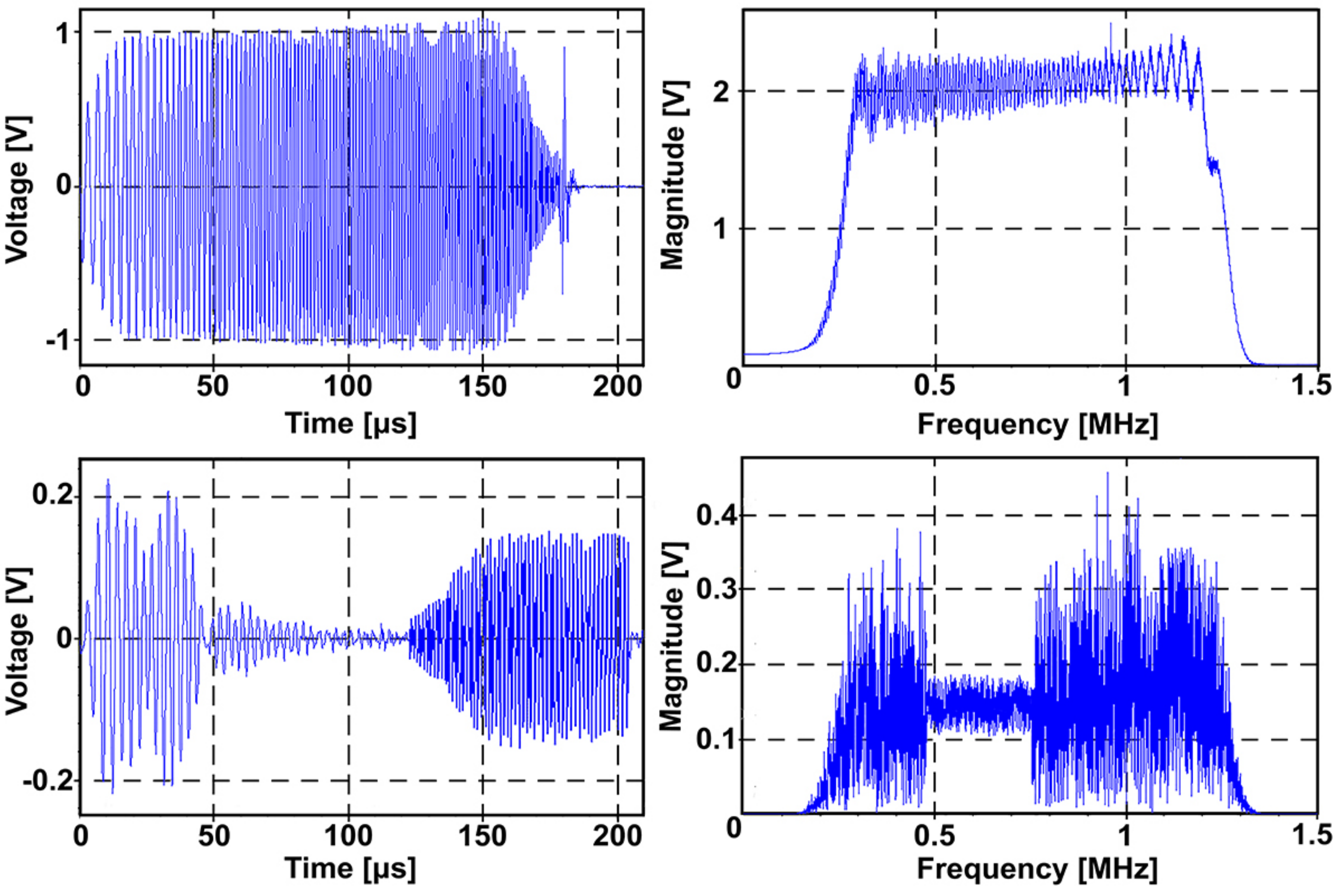}
  \caption{Experimental results for a {\it bcc} PnC slab with 8 layers and steel spheres of $2\,mm$ diameter in paraffin. The frequency bandgap extends between 480 and 750 KHz.}
  \label{fig15}
\end{figure}

\begin{figure}
  \centering
  \includegraphics[width=0.7 \textwidth]{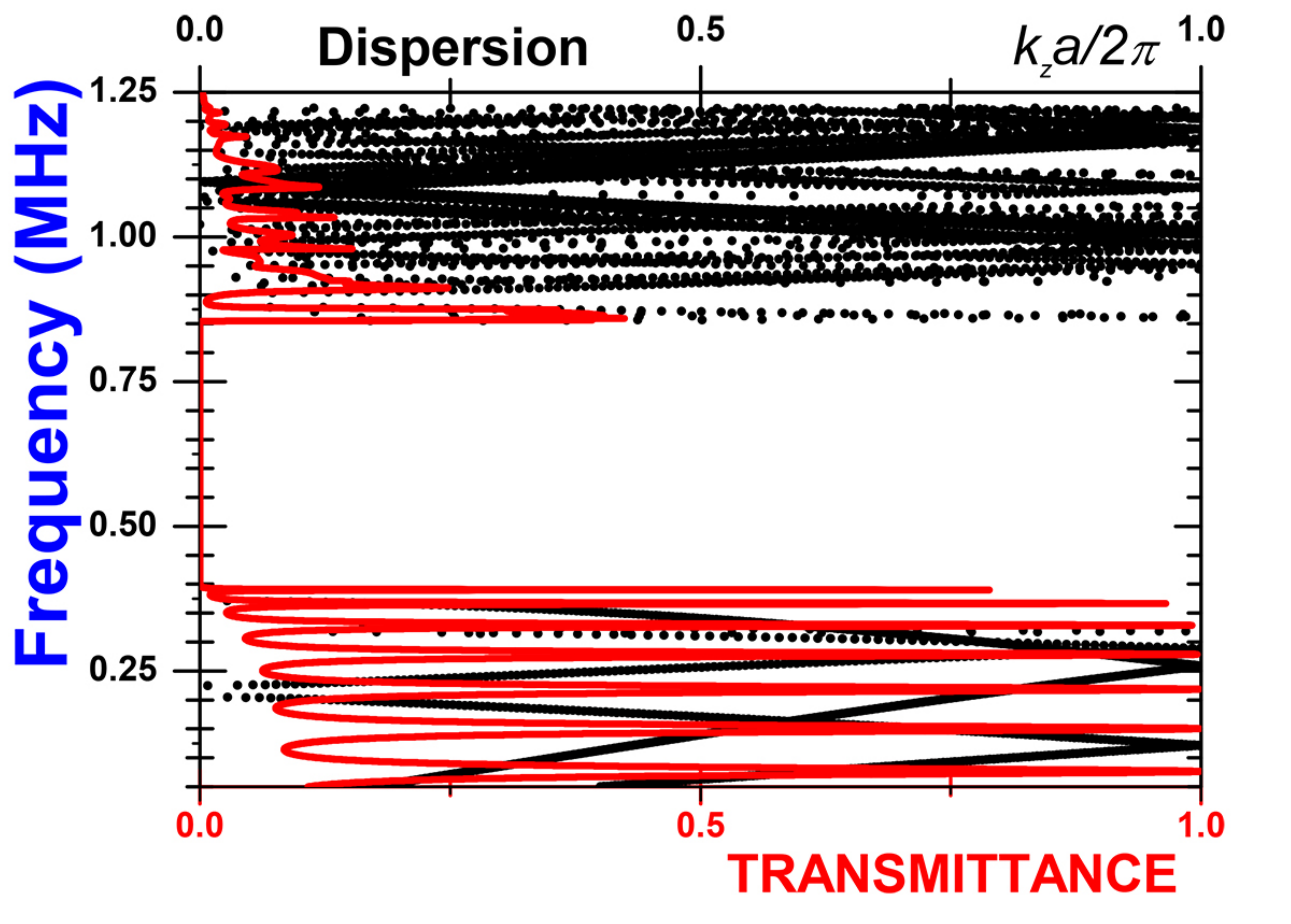}
  \caption{Frequency band structure at the center $\mathbf{k}_{\parallel}= \mathbf{0}$ of the Surface Brillouin Zone (SBZ) of the (001) surface of an {\it hcp} PnC (denoted as dispersion).
  The crystal consists of $2\,mm$ in diameter stainless steel balls in paraffin. The transmission spectrum of a longitudinal wave along the same direction from a slab of the crystal 8 layers thick is presented in red.}
  \label{fig16}
\end{figure}

\begin{figure}
  \centering
  \includegraphics[width=0.7 \textwidth]{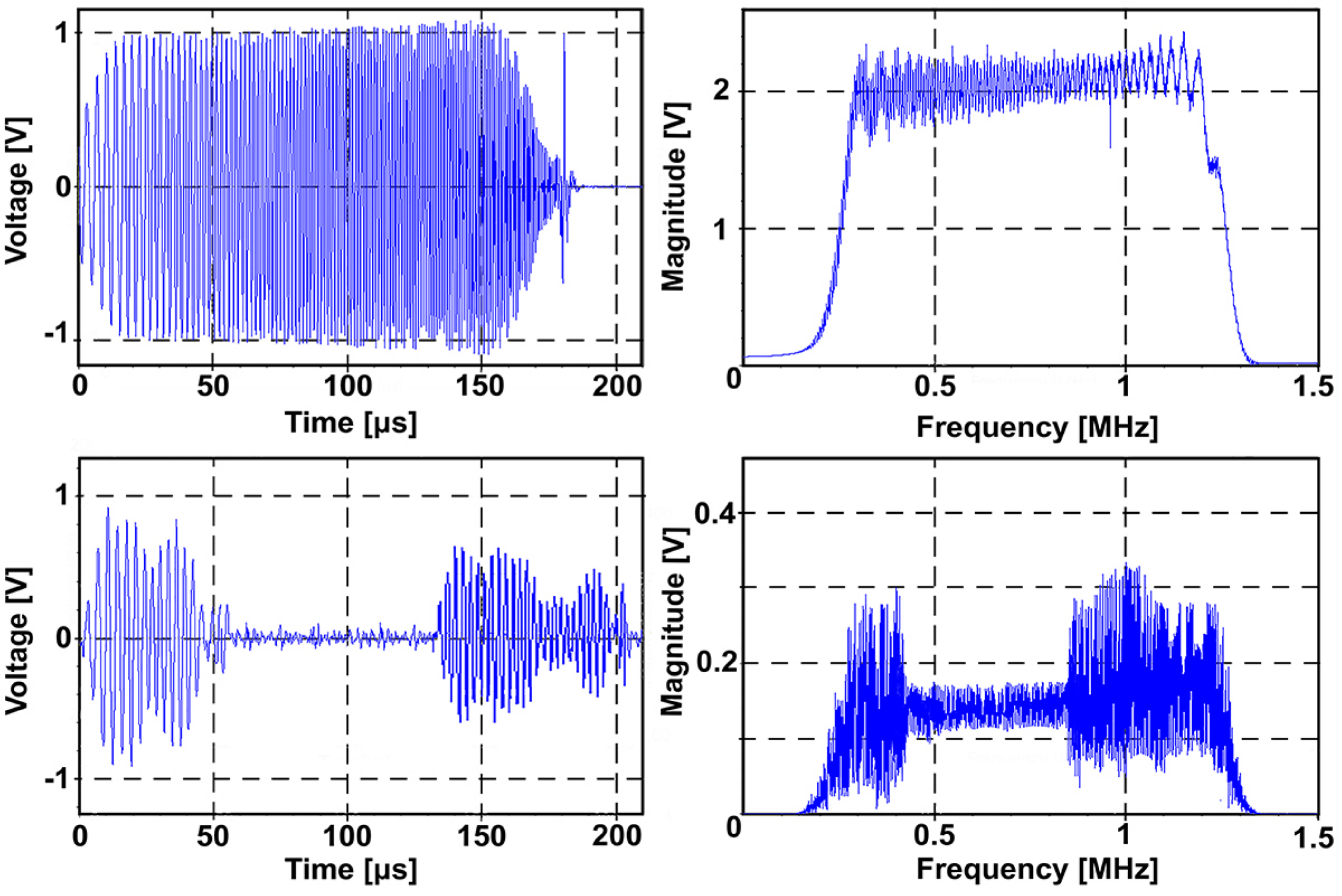}
  \caption{Experimental results of {\it hcp} PnC slabs with 8 layers and steel spheres of $2\,mm$ diameter in paraffin.}
  \label{fig17}
\end{figure}

\begin{figure}
  \centering
  \includegraphics[width=0.7 \textwidth]{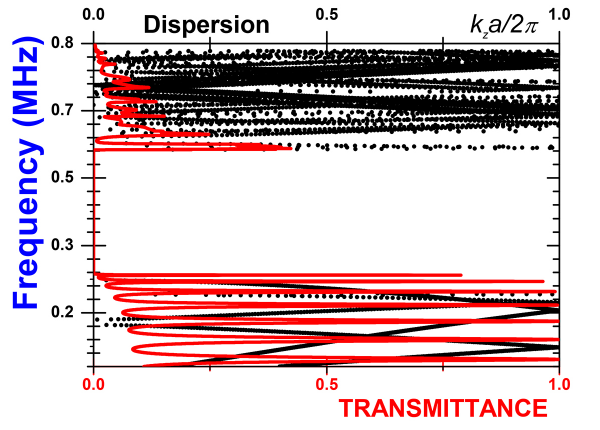}
  \caption{Dispersion and transmission spectrum of the same {\it hcp} PnC and slab as in Fig.~\ref{fig16}, along the same crystallographic direction. In this case the spheres are $3\,mm$ in diameter.}
  \label{fig18}
\end{figure}

\begin{figure}
  \centering
  \includegraphics[width=0.7 \textwidth]{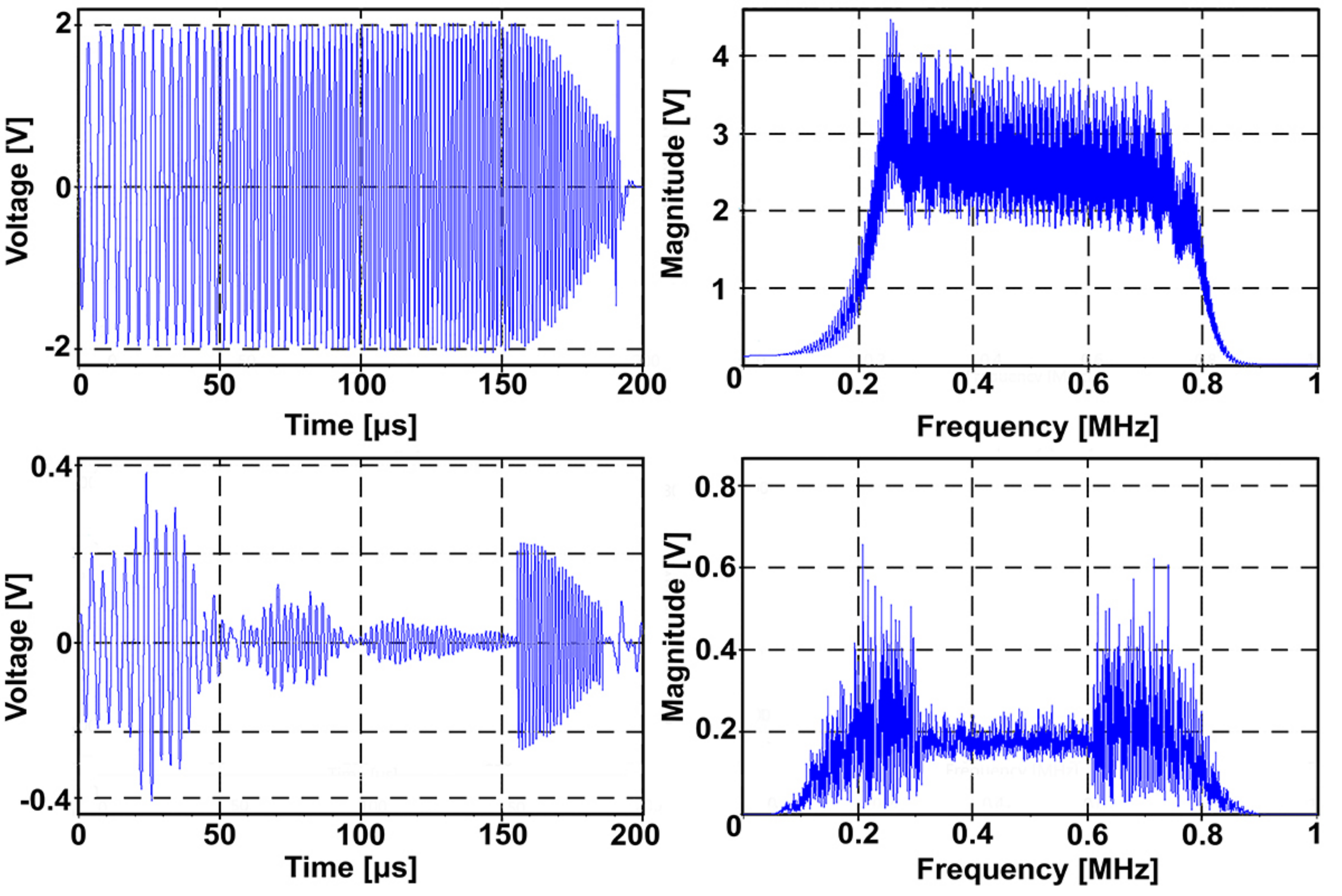}
  \caption{Experimental results for a PnC slab, 8 layers thick, following the specifics presented in Fig.~\ref{fig18}.}
  \label{fig19}
\end{figure}

\begin{figure}
  \centering
  \includegraphics[width=0.7 \textwidth]{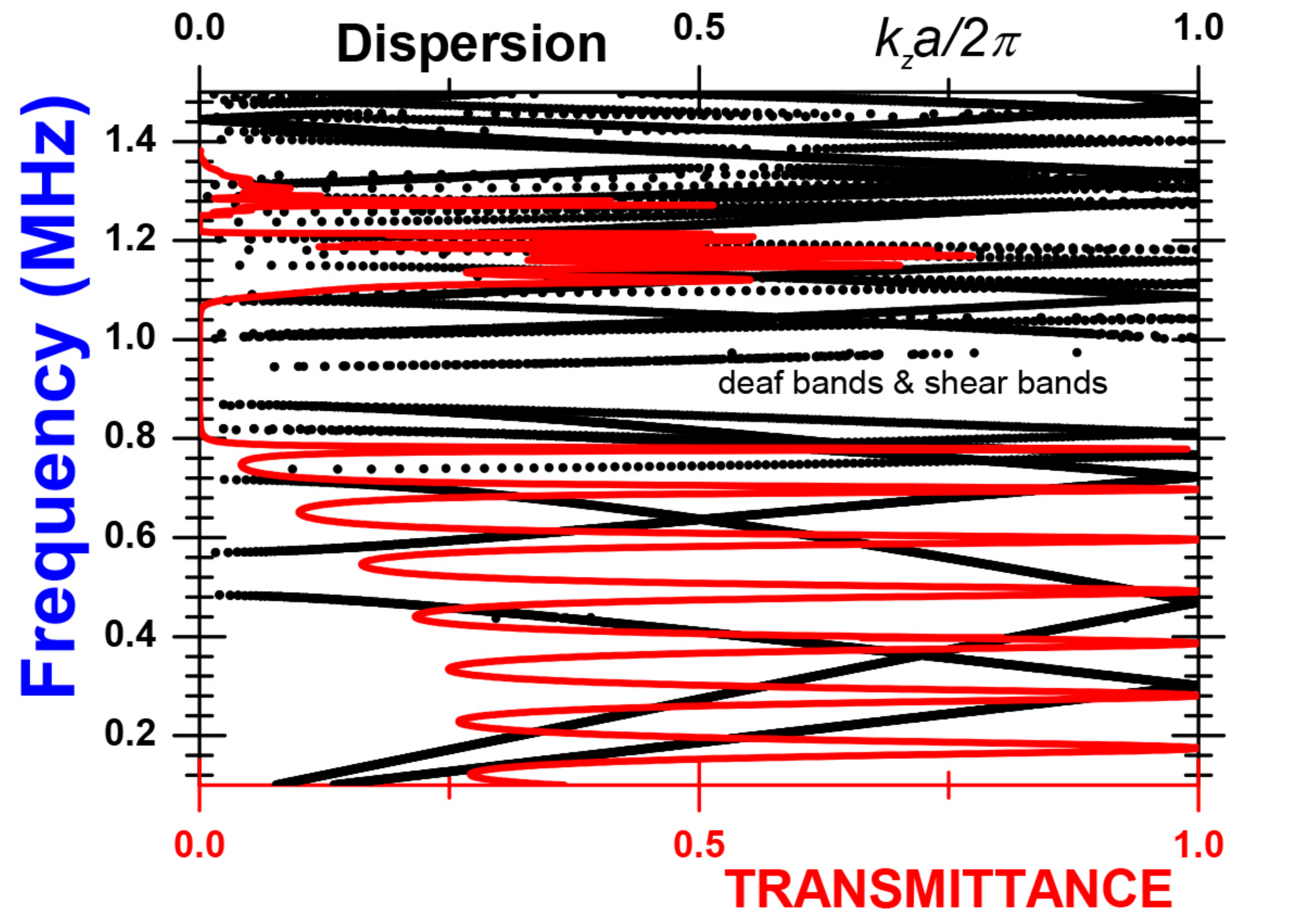}
  \caption{Dispersion and transmission response of an {\it hcp} PnC and a slab 8 layers thick, consisting of $3mm$ steel spheres in cement paste, along the [001] direction.
  Inside the gap deaf and shear bands remain inactive for longitudinal normal incidence.}
  \label{fig20}
\end{figure}

\begin{figure}
  \centering
  \includegraphics[width=0.7 \textwidth]{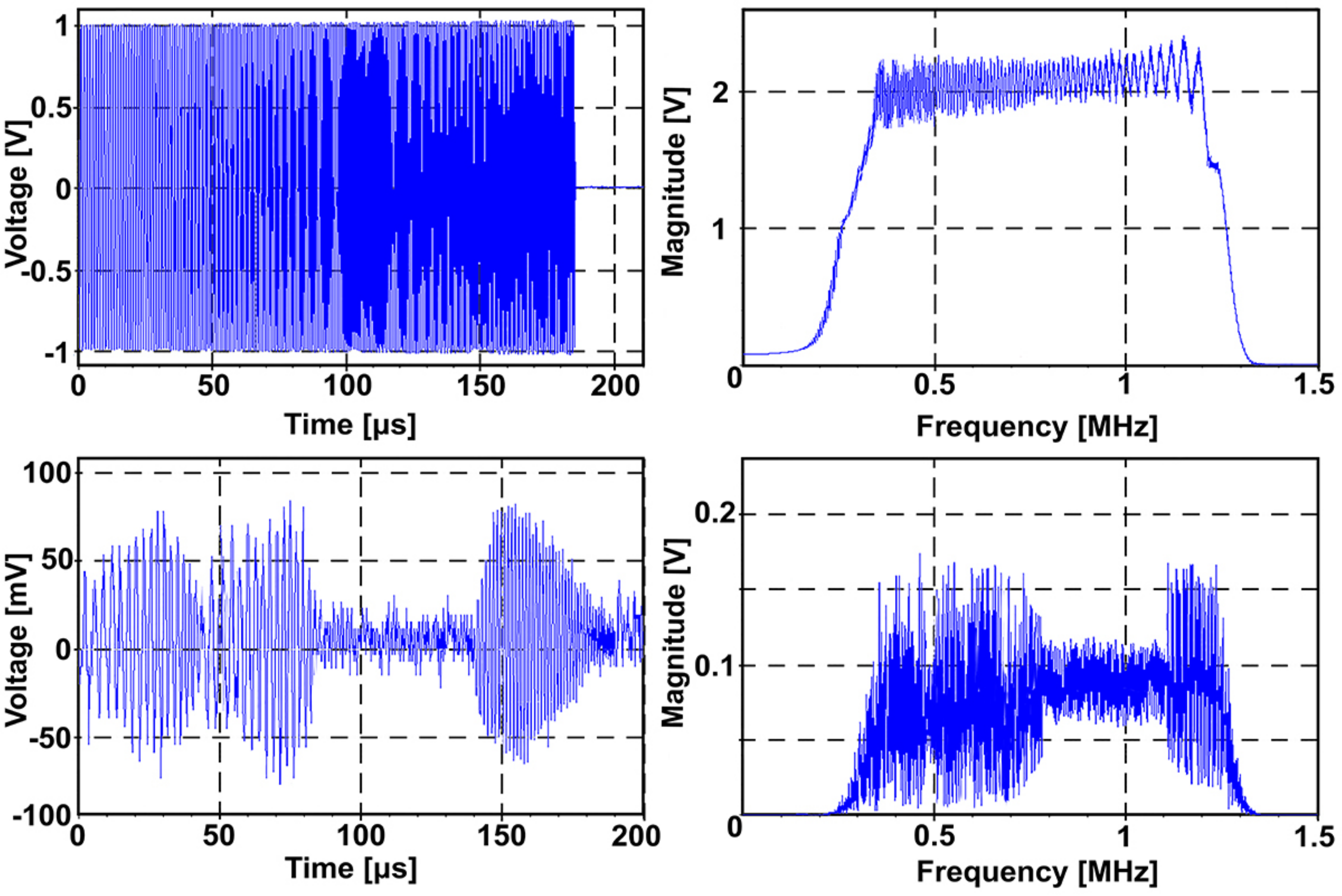}
  \caption{. Experimental results for 8 layers thick {\it hcp} PnC slab consisting of $3mm$ in diameter steel spheres embedded in cement paste matrix.}
  \label{fig21}
\end{figure}

\section{Experimental investigation of PnC slabs}
For the phononic gaps experiments, a high voltage pulse generator RITEC  (RPR-4000) able to reproduce pulses of frequencies ranging from 0 to 22 MHz, up to $1000V$ peak-to-peak, was used. As the generator cannot produce chirp signals, a Tektronic TDS 1012B pulse generator was used as amplifier. In addition, a 2D laser Doppler vibrometer (LDV), Polytec PCV-400, was used as a non-contact sensor for performing non-contact vibration measurements of the surface of the samples with the aid of a laser beam. Vibration amplitudes and frequencies are extracted from the Doppler shift of the reflected laser beam frequency due to the motion of the surface. LDV is advantageous in that the laser beam can be directed at the surface of interest as well as in that the vibration measurement does not impose extra weight loading on the target structure. The technique has been used as a vibration sensor in aerospace, industry, research, for crack detection in metallic structures, civil and
mechanical engineering industries~\cite{LDVibe,Staszw,Staszw2,Nassif,Khalil,Watt,Stan}. Contact piezoelectric crystal-based ultrasound sensors with central frequency of 200 KHz – 1.5 MHz and spectrums distributed around its maximum frequency were also used.To guide ultrasound waves on the surface of the PnC slab, an aluminum waveguide of dimensions of $150 \times 100 \times 100\, mm^3$, with an additional central cube of dimensions $10 \times 10 \times 10\, mm^3$ in the center of one surface, was used (Fig.~\ref{fig07}).

Figure~\ref{fig10} shows the experimental arrangement for lock-in IR thermography, comprising of the IR camera, the lamps, the pulse generator for thermal stimulation of the PnC, and the processing unit of the thermographic results.

\section{Results and Discussion}
The vibrated stimulation of the PnCs was performed using a chirp signal, i.e. a constant amplitude signal with increased frequency over time. The phononic slabs were subjected to a range of vibration frequencies and their response to these frequency ranges was measured.  This way, the stop bands  for a specific frequency range was found. Initially, a pure paraffin structure with the same thickness as the PnC slabs, without stainless steel spheres, was evaluated. Figure~\ref{fig11} shows the resulting spectrum of application of a chirp vibration in pure paraffin, for a wide range of frequencies from 200 KHz to 1.25 MHz. As observed in this figure, the spectrum in the entire frequency range is continuous and does not display frequency gaps.

For the design, construction  and evaluation of materials that will provide shields to the vibrations, the physics of the interaction of elastic waves was studied. LMS is the most reliable and complete solution for modeling PnCs in 3 dimensions. It is also the theoretical approach closer to a real experiment, since it calculates the coefficient of passage of acoustic (elastic) waves from a finite tile of the crystal. In the following, theoretical  results based on the LMS method are presented and compared to experimental data . First, the effect of application of a chirp signal (750 KHz - 4 MHz) with a 1.2 – 3.2 MHz high-precision filter is depicted in Fig.~\ref{fig12}. In this figure, the two upper frames show the reference chirp signal on time domain and FFT while the two lower frames are the results of cut filter. The chirp signal is sent in bursts of duration of 200 $\mu s$, so that each frequency is generated in a specific amount of  time.

Figure~\ref{fig13} shows the results for pure paraffin. Therein, no frequency gap is observed. Figure~\ref{fig14}) shows the LMS-expected frequency gaps in a 8-layer thick {\it bcc} phononic slab with $2\, mm$ diameter spheres as they appear in the appropriate transmission spectrum. In addition, an extra calculation was performed to account for all measured losses. It is evident that the positions of silent zones are not affected by any losses present in the system. This type of behavior was expected, following the theoretical predictions presented in Ref.~\cite{ipsa02}. Figure~\ref{fig15} shows the experimental result of a slab of a {\it bcc} PnC with 8 layers and $2\, mm$ diameter balls. The experimental results for this case are in full agreement  with the  results of LMS method. In particular, a frequency gap in the range of 480 - 750 KHz was observed.

Figure~\ref{fig16} shows the LMS results (directional frequency band structure of an infinite crystal) of a {\it  hcp} PnC with 8 layers and $2\,mm$ spheres while Fig.~\ref{fig17} shows the experimental results of the same. The observed gap is in the range of  420 -850 KHz as it is supported by theory. In addition, the {\it hcp} arrangement exhibits larger gap as compared to its {\it bcc} counterpart. The red lines in Fig.~\ref{fig16} show the fluctuation of a longitudinal wave incident perpendicular to a tile of the crystal.

The frequency of gaps are reduced significantly for the $3\,mm$ inclusions. By examination of Figures ~\ref{fig18} and ~\ref{fig19}, which represent theory and experiment, respectively, frequency gap appears between 290 and 600 KHz while a range of 280 - 590 KHz is theoretically predicted.

The results of a PnC with the cement paste matrix are given in Figures~\ref{fig20} and ~\ref{fig21} and feature an {\it hcp} phononic slab, 8 layers thick with steel spheres of $3\, mm$ in diameter. In this case, the frequency gap appears in the range of 0.8 - 1.1 MHz. Agreement of theory and experiment follows as observed in all cases so far. In particular, it is interesting to mention that that the presence of deaf and shear bands in the band structure of Fig~\ref{fig20} do not couple with the external elastic field and therefore remain inactive. Fact which is clearly seen on the theoretical transmission spectrum and the experimental results on Fig.~\ref{fig21}.

A comparative summary of all theoretically determined vs experimentally obtained frequency gap ranges, widths and gap width over midgap frequency is presented in Table 2. Therein, one can clearly see that the measured observables are consistent with the underlying theory.

\begin{table}[t]
  \centering
    \caption{Comparison between theory and experiment. $\Delta \omega / \omega_G$ represents the gap width over midgap frequency and denotes the gap efficiency, which can be scaled to a wide range of size parameters.}
   \label{tab:table2}
    \begin{tabular}{l|c|c|c|c|c|c} 
    \hline
    \hline
      PnC type & Freq. Gap & Freq. Gap & Gap Width & Gap Width & $\Delta \omega / \omega_G$ & $\Delta \omega / \omega_G$\\
      (sphere diameter)  & Theory  & Exp. & Theory  & Exp. & Theory  & Exp.\\
      \hline
      \  & KHz & KHz & KHz & KHz & \% & \% \\
      \hline
      \hline
      Paraffin $hcp$ ($2mm$) & 400-850 & 420-850 & 450 & 430 & 72 & 68\\
      Paraffin $hcp$ ($3mm$) & 280-580 & 280-590 & 300 & 310 & 69.8 & 71.9\\
      Paraffin $bcc$ ($2mm$) & 450-750 & 460-750 & 300 & 290 & 50 & 48\\
      Cement $hcp$ ($3mm$) & 800-1100 & 820-1050 & 300 & 230 & 31.6 & 25\\
      \hline
      \hline
    \end{tabular}
\end{table}

\section{Conclusion}
In the present study, a nondestructive technique based on LDV was employed in order to record the band-gap formation of 3D PnCs. Several types of PnC slabs of varying crystallographic symmetry in paraffin and cement matrices were fabricated and their frequency gap spectrum was recorded experimentally using the LDV method. The results were compared with theoretical expectations and good agreement was found in all cases. The favorable comparison renders the LDV technique an effective and promising tool for the non-contact, non-destructive assessment of PnCs. The developed methodology has the potential to be equally versatile in cases where classical evaluation methods, requiring contact sensors, is not possible.  Such cases may include very small regions, high temperature  conditions  as well as nanostructured  metamaterials, where the mass of a contact sensor can affect the experimental results. Finally, it should be noted that the experimental methodology developed in this study does not impose any size or scale limitations concerning the evaluation of a PnC slab.

\section*{Acknowledements}
This research has been co-financed by the European Union (European Regional Development
Fund- ERDF) and Greek national funds through the Operational Program Education and Lifelong
Learning 2007-2013 of the National Strategic Reference Framework (NSRF 2007-2013), Action
ARISTEIA II.



\bibliography{psarobas_bib}

\end{document}